\documentclass[conference]{IEEEtran}
\IEEEoverridecommandlockouts
\usepackage{listings}

\newcommand*\rotV{\rotatebox{90}}

\usepackage{multirow}

\usepackage{balance}
\usepackage{flushend}
\usepackage{cite}
\usepackage{amsmath,amssymb,amsfonts}
\usepackage{algorithmic}
\usepackage{graphicx}
\usepackage{textcomp}
\usepackage{url}
\usepackage{hyperref}
\usepackage{multirow}
\usepackage{subcaption}

\usepackage{xcolor}
 \usepackage{enumitem}
\def\BibTeX{{\rm B\kern-.05em{\sc i\kern-.025em b}\kern-.08em
    T\kern-.1667em\lower.7ex\hbox{E}\ kern-.125emX}}
    
\begin{document}

\title{Using Microservice Telemetry Data for System Dynamic Analysis
\thanks{This material is based upon work supported by the National Science Foundation under Grant No. 1854049, grant from Red Hat Research, and Ulla Tuominen (Shapit).}
}

\author{
\IEEEauthorblockN{1\textsuperscript{st} Abdullah Al Maruf}
\IEEEauthorblockA{\textit{Department of Computer Science} \\
\textit{Baylor University}\\
Waco, Texas, United States \\
maruf\_maruf1@baylor.edu}
\and
\IEEEauthorblockN{2\textsuperscript{nd} Alexander Bakhtin}
\IEEEauthorblockA{\textit{Software Engineering Group} \\
\textit{Tampere University}\\
Tampere, FI-33720, Finland \\
alexander.bakhtin@tuni.fi}
\and
\IEEEauthorblockN{3\textsuperscript{rd} Tomas Cerny}
\IEEEauthorblockA{\textit{Department of Computer Science} \\
\textit{Baylor University}\\
Waco, Texas, United States \\
Tomas\_Cerny@baylor.edu}
\and
\IEEEauthorblockN{4\textsuperscript{th} Davide Taibi}
\IEEEauthorblockA{\hspace{22em}\textit{Software Engineering Group\hspace{22em}} \\
\textit{Tampere University}\\
Tampere, FI-33720, Finland \\
davide.taibi@tuni.fi}
}

\maketitle

\begin{abstract}
Microservices bring various benefits to software systems. They also bring decentralization and lose coupling across self-contained system parts. Since these systems likely evolve in a decentralized manner, they need to be monitored to identify when possibly poorly designed extensions deteriorate the overall system quality. For monolith systems, such tasks have been commonly addressed through static analysis. However, given the decentralization and possible language diversity across microservices, static analysis tools are lacking. On the other hand, there are available tools commonly used by practitioners that offer centralized logging, tracing, and metric collection for microservices. In this paper, we assess the opportunity to combine current dynamic analysis tools with anomaly detection in the form of quality metrics and anti-patterns. We develop a tool prototype that we use to assess a large microservice system benchmark demonstrating the feasibility and potential of such an approach.


\end{abstract}


\section{Introduction} \label{sec:introduction}

Microservice architecture (MSA) has become the industry standard. MSA allows for efficient scaling, improved resiliency, and faster software delivery because of its loosely connected nature. It also enables teams to focus on a single task/module and produce reliable software faster by allowing them to use any programming language that suits the task.
With the ability of agility and speed of development, MSA needs constant monitoring and analysis to tackle the complexity of the system. A microservice system's longevity and quality can be jeopardized without regular monitoring.

The system can be subjected to static and/or runtime analysis. While static analysis can detect bugs, vulnerabilities, and smells \cite{walker2020smells,8906700} before putting software into production, it also has the disadvantage of requiring language-specific analysis. It is challenging to find language-specific static analyzer tools unless the modules are written in a few popular languages. Otherwise, the developers need to create their tool, which will increase maintenance workload, or they must exclude those modules from analysis, which is also not a good option. Runtime analysis can help solve this language-specific issue. Telemetry data is one form of dynamic analysis which provides a few key analysis perspectives, such as Service Dependency Graph (SDG), detecting architectural smells, and architectural evolution using the SDG.

This paper uses telemetry data to determine inter-service communication and build the Software Dependency Graph. Next, it uses this information to find architecture smells, calculate anti-pattern metrics automatically, detect architectural evolution, and effectively scale the microservice modules. This can help system operators and engineers better understand the system specifics and concerns that should be addressed throughout the system evolution.

The following is a breakdown of the paper's structure: Section \ref{sec:background} discusses static and dynamic analysis and prior studies on the analysis of microservice systems. Then Section \ref{sec:applications} goes over various types of telemetry data and their applications. Section \ref{proposal} proposes a method for generating SDGs from telemetry logs. We discuss possible analysis directions in Section \ref{analysis} and showcase a case study in 
Section \ref{case-study}. We conclude the paper with Section \ref{conclusion}.

\section{Background \& Related Work} \label{sec:background}


The static analysis uses software artifacts such as source code, deployment manifests, and API documentation. Dynamic analysis is the real-time testing or profiling of a system. Dynamic analysis can be applied to runtime data from a system in either a production or a staging/development environment. Users must use the system, or a script must replicate real-time user behavior, such as accessing all use cases and making reasonable user requests per minute, to generate runtime data. Data obtained in runtime includes application logs and telemetry data. Both these approaches have their specifics and suitability for certain tasks. However, certain overlaps exist, such as the aspects we address in this paper.

Detecting bad smells and poor design indicators in a decentralized environment have been addressed in the literature. For instance, Taibi et al. \cite{Taibi2018IEEE} identified recurrent smells in microservices. There has been approached by Walker et al. \cite{walker2020smells}, or Pigazzini et al. \cite{Pigazzini2020} in order to detect them. Both these approaches involved static analysis or a combination of static and dynamic analysis, which is currently limited to a specific platform. The major challenge is to reconstruct the holistic system perspective. A perspective that makes it obvious for practitioners to understand how the system divides into specific microservices and how these interact and depend on one another. Rademacher et al. \cite{10.1007/978-3-030-49418-6_21}, Walker et al. \cite{walker2021automatic} or Granchelli \cite{7958455} considered the process software architecture reconstruction for microservices. While it can be performed manually from various artifacts \cite{10.1007/978-3-030-49418-6_21}, combining static and dynamic analysis \cite{7958455} or an automated static analysis approach using distributed codebase is also viable \cite{walker2021automatic} as demonstrated by Bushong et al. \cite{closer22}. The major challenge with static analysis is the code analysis language dependency and distribution. While the initial approach to address this challenge has been proposed \cite{schiewe2022}, the tooling support is not yet available.

The tooling support is broader in the realm of dynamic analysis. Common tools exist for centralized logging, distributed tracing, and telemetry. In addition, various system-centric perspectives are available in such tools (i.e., Jaeger, Kiali, etc.), and it is possible to preview trace-reconstructed system topology or dependency graphs. While these views cannot be as detailed as when assessing the code, it gives sufficient abstraction on the running system and language agnosticism. Given the broad availability of these tools, it is reasonable to consider the integration of detection of various patterns~\cite{Taibi2018closer18},  anti-patterns~\cite{Taibi2020MSE}, smells~\cite{Taibi2018IEEE}, or quality metrics (e.g. Coupling~\cite{Panichella2021}) using dynamic analysis. While it might be assumed that development and operations (DevOps) engineers can easily see these indicators, Bento et al. \cite{dynamic2021} suggested that the current tools do not provide appropriate ways to abstract, navigate, filter, and analyze tracing data and do not automate or aid with trace analysis. Instead, the process relies on DevOps, but these might lack the expertise or the time necessary to determine the statistics in the ever-changing environment.

When using traces, it can be seen that SDG is commonly used. For instance, Ma et al. \cite{8377834} use it to analyze and test microservices through graph-based microservice analysis and testing. However, Ma et al. made the process dependent on DevOps manual efforts to detect anomalies by analyzing risky service invocation chains and tracing the linkages between services.

In this paper, we look into the automation of quality metric detection using dynamic analysis of telemetry data, which would indicate the areas of concern for DevOps and let them prioritize their tasks.




\section{Telemetry Data \& Its Applications} \label{sec:applications}

In software engineering, telemetry data refers to collecting data from software and systems that indicate the source's state. Telemetry is one of the key factors in increasing the observability of a complex system \cite{karumuri2021towards}. Johnson et al. \cite{johnson2005improving} defined software project telemetry as a method of defining, collecting, and analyzing software metrics that has the five characteristics listed below:

\begin{itemize}
    \item Data should be collected automatically, with no manual intervention from humans.
    \item A timestamp must be assigned to each event in the data.
    \item Every project member has continuous and immediate access to the data.
    \item Telemetry analysis should be valuable even if it lacks complete data for the entire project's lifespan.
    \item Telemetry analyses depict the project's current state and how it evolves over time.
\end{itemize}

Opentelemetry \cite{opentelemetry.io_2020} is the standardization of the telemetry data collection process. It is difficult for developers to switch tools and adapt to a new tool because of the different standardization prior to Opentelemetry.  Opentelemetry provides a vendor-agnostic API for sending telemetry data to a backend and a set of language-specific libraries for instrumenting code and shipping data to one of the supported backends. 

Telemetry data can be categorized into three main formats  \cite{karumuri2021towards,picoreti2018multilevel,opentelemetry.io_2020}: 
\begin{itemize}
     \setlength{\itemindent}{4em}
    \item Traces
    \item Metrics, and
    \item Logs.
\end{itemize}

Karumuri et al. defined metrics as numeric data, logs as unstructured strings, and traces as a graph of a request's execution path \cite{karumuri2021towards}. Along with the available libraries, developers can add software-specific metrics to be shipped to the backend, such as Prometheus. For example, if there are many unsent emails in the queue. In that case, developers can ship the number of unsent emails to Prometheus and then use Alertmanager to trigger an alert. Grafana is a popular tool for visualizing metrics, and developers can use a shared grafana dashboard, a platform for sharing custom dashboards that use critical metrics.

There are a few tools that can be used to trace a request in order to analyze it. Some examples of such tools are Zipkin and Jaeger. We can use tracing to find the source of slow response and keep track of distributed transactions. The majority of tracing tools also provide a software architecture graph in the form of a service dependency graph.

Although there are a few tools for tracing, metrics analysis, and visualization, there is little for log analysis. Most developers use ELK (Elasticsearch, Logstash, and Kibana) to store and process a distributed system's log centrally. Elasticsearch stores large amounts of data while also allowing for faster data searches. Logstash serves as a log aggregator and processor, reading logs and sending them to Elasticsearch. Kibana allows users to run queries against log data. It is common to combine tracing and logs so that developers can search for error logs with the request's traceID across the entire distributed system.

Even though telemetry requires a language-specific SDK library to produce metrics or traces, service mesh adds this capability without requiring the addition of a library to the source code. This eliminates the need for source code changes and the language barrier. In addition to tracing, service mesh includes metrics and access logs by default.

\section{Service Mesh \& Service Dependency Graph (SDG)} \label{proposal}

\begin{figure}[th]
  \includegraphics[width=.5\textwidth]{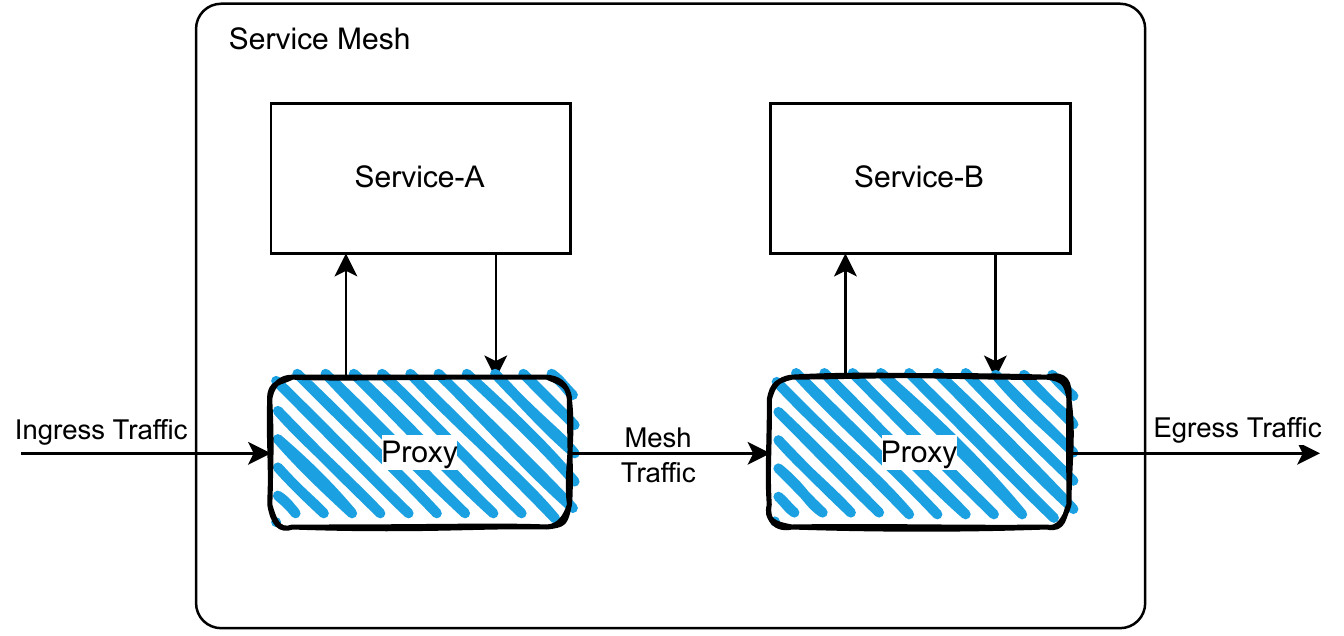}
  \caption{Architecture of a service mesh where proxy sidecars handles all the network activity of a service.}
  \label{fig:istio-arch}
\end{figure}

Distributed systems are notoriously difficult to manage, and adding traffic management to the mix makes things even more difficult. A service mesh comes with traffic management, observability, security, encryption, access control, rate limiting, and other features out of the box. As a result, service mesh has become an essential tool in Kubernetes-based systems. Each service module is deployed separately in this system, and each of these modules has a sidecar \cite{burns2016design} proxy that receives and sends traffic on behalf of the microservice module (shown in figure-\ref{fig:istio-arch}). These proxy sidecars can control traffic and improve the service's observability and security.

Istio \cite{istio.io_2020}, linkerd\footnote{https://linkerd.io/}, and consul connect\footnote{https://www.consul.io/docs/connect} are the most popular open source service mesh tools, according to GitHub stargazers. In addition to these tools, most service mesh tools provide an access log (also known as audit logging or proxy log), containing information about each inbound and outbound request in a customizable format. Our method uses these access logs to determine which service is calling which service, which endpoint is being called, and how frequently a service is called in a given time period. We can use this information to reconstruct software architecture and automatically detect anti-patterns and coupling.

\lstset{basicstyle=\footnotesize\ttfamily,breaklines=true}
\begin{lstlisting}[
  caption={Default Format of Access Logs in Istio service mesh}
  \label{lst:format},
]
[%START_TIME%] \"%REQ(:METHOD)% %REQ(X-ENVOY-ORIGINAL-PATH?:PATH)% %PROTOCOL%\" %RESPONSE_CODE% %RESPONSE_FLAGS% %RESPONSE_CODE_DETAILS% %CONNECTION_TERMINATION_DETAILS%
\"%UPSTREAM_TRANSPORT_FAILURE_REASON%\" %BYTES_RECEIVED% %BYTES_SENT% %DURATION% %RESP(X-ENVOY-UPSTREAM-SERVICE-TIME)% \"%REQ(X-FORWARDED-FOR)%\" \"%REQ(USER-AGENT)%\" \"%REQ(X-REQUEST-ID)%\"
\"%REQ(:AUTHORITY)%\" \"%UPSTREAM_HOST%\" %UPSTREAM_CLUSTER% %UPSTREAM_LOCAL_ADDRESS% %DOWNSTREAM_LOCAL_ADDRESS% %DOWNSTREAM_REMOTE_ADDRESS% %REQUESTED_SERVER_NAME% %ROUTE_NAME%\n
\end{lstlisting}

\begin{table*}[]
\centering
\begin{tabular}{|l|l|}
\hline
\multicolumn{1}{|c|}{Request Type} & \multicolumn{1}{c|}{Sample Access Log} \\ \hline
inbound request                    & \begin{lstlisting}
{
  "start_time": "2022-05-26T06:22:02.661Z",
  "upstream_host": "10.244.0.65:12345",
  "downstream_local_address": "10.96.162.171:12345",
  "upstream_transport_failure_reason": null,
  "protocol": "HTTP/1.1",
  "upstream_service_time": "6",
  "authority": "b-service:12345",
  "requested_server_name": null,
  "response_code_details": "via_upstream",
  "connection_termination_details": null,
  "upstream_local_address": "10.244.0.41:33326",
  "downstream_remote_address": "10.244.0.41:48250",
  "path": "/api/v1/endpoint/",
  "bytes_sent": 44,
  "request_id": "4631dc4c-0a6e-9ad2-ba61-d257cdd6e50b",
  "bytes_received": 0,
  "route_name": "default",
  "duration": 7,
  "x_forwarded_for": null,
  "response_flags": "-",
  "response_code": 200,
  "method": "GET",
  "upstream_cluster": "outbound|12345||b-service.default.svc.cluster.local",
  "user_agent": "Apache-HttpClient/4.5.9 (Java/1.8.0_111)"
}
\end{lstlisting}\\ \hline
outbound request                   & \begin{lstlisting}
{
  "start_time": "2022-05-26T06:22:02.661Z",
  "upstream_host": "10.244.0.65:12345",
  "downstream_local_address": "10.96.162.171:12345",
  "upstream_transport_failure_reason": null,
  "protocol": "HTTP/1.1",
  "upstream_service_time": "6",
  "authority": "b-service:12345",
  "requested_server_name": null,
  "response_code_details": "via_upstream",
  "connection_termination_details": null,
  "upstream_local_address": "10.244.0.41:33326",
  "downstream_remote_address": "10.244.0.41:48250",
  "path": "/api/v1/endpoint/",
  "bytes_sent": 44,
  "request_id": "4631dc4c-0a6e-9ad2-ba61-d257cdd6e50b",
  "bytes_received": 0,
  "route_name": "default",
  "duration": 7,
  "x_forwarded_for": null,
  "response_flags": "-",
  "response_code": 200,
  "method": "GET",
  "upstream_cluster": "outbound|12345||b-service.default.svc.cluster.local",
  "user_agent": "Apache-HttpClient/4.5.9 (Java/1.8.0_111)"
}
\end{lstlisting}\\ \hline
\end{tabular}
\caption{An example of an inbound and outbound request access log. Inbound requests have a-service as their destination, while outbound requests have a-service as their source and b-service as their destination.}
\label{tab:access-log}
\end{table*}

\begin{figure}[th]
  \includegraphics[width=.45\textwidth]{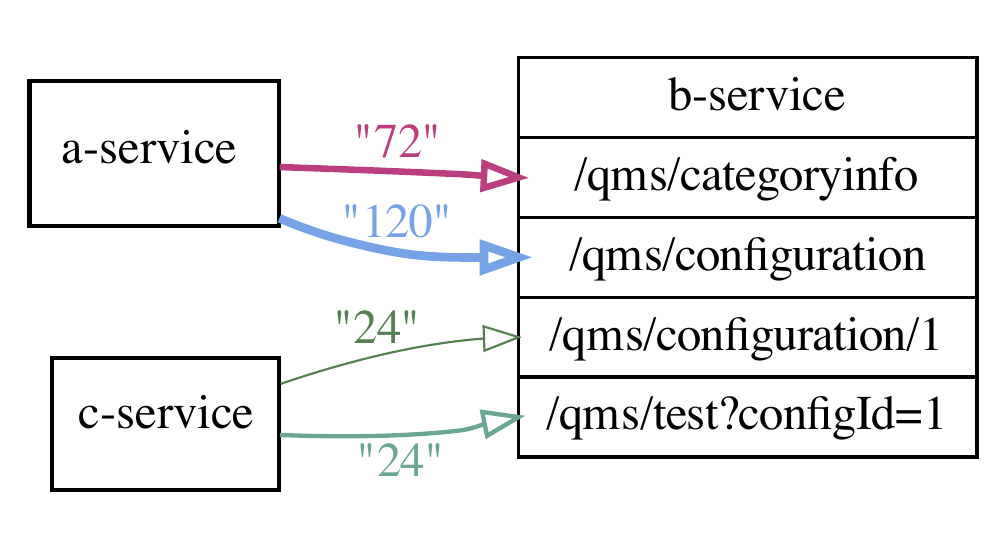}
  \caption{A SDP prototype that can be generated from an access log. Each edge weight number represents the number of times the endpoint is called. When the weight value is greater, the edge thickens.}
  \label{fig:sample-sdg}
\end{figure}

The default format of the access logs can be seen in the Listing \ref{lst:format}. The Envoy proxy website lists the definitions of each term of this format, as well as other customizable formats \cite{envoyauthors_2020}. Table \ref{tab:access-log} contains a sample inbound and outbound request’s ‘access-log’, where both logs are from service-a's proxy sidecar, and its service DNS inside the Kubernetes cluster is `a-service.default.svc.cluster.local'. One thing to note is that while both the inbound and outbound logs contain downstream and upstream addresses in the form of IPs, IPs are ephemeral in Kubernetes-based systems because they rely on ephemeral `Pods' \cite{marmol2015networking}. The service's DNS, on the other hand, is stable, and it will forward traffic to a Pod's or a group of pods' updated IP address \cite{marmol2015networking}.

The outbound request indicates that the event log contains upstream service information in DNS format, with the key `upstream cluster' and `path' indicating which upstream service endpoint is being requested. We can use this information to construct the SDG, in which a-service and b-service each represent a node, with a directed edge connecting a-service to b-service's endpoint \texttt{/api/v1/endpoint/}. We can also maintain the interaction count as edge-weight on this directed edge.' 

We can analyze each access log of a distributed system and generate an edge in the graph for each interaction between two services within a time period. If the services already have an edge, we can increase the weight of the edge by one. Finally, we can use a visualizing tool to draw the graph automatically. We can make the edge visually thinner or thicker depending on the edge weight. Using different colors for different edges is also effective. Figure \ref{fig:sample-sdg} shows a prototype of such an SDG, in which three services are represented by three nodes called `a-service,' `b-service,' and `c-service,' and each edge represents a request from a source service to a destination service endpoint.

\section{Possible Analysis Directions} \label{analysis}

Using runtime logs to generate SDG and obtain service connectivity information eliminates the need for a static method's language-specific analyzer. We can use this information to detect anti-patterns and the evolution of the system architecture after collecting communication information between services and generating an SDG. The criticality and reliability metrics of the microservice architecture (MSA) \cite{rud2006product} can also be measured automatically.

\subsection{Architecture smell/Anti-Pattern detection}

Architectural smells are one of the causes of architectural decay, and technical debt \cite{le2015empirical, fellah2019architectural}. Finding architectural smells and anti-patters can thus aid the software's long-term viability. We compiled a list of anti-patterns from Borges et al. \cite{borges2019algorithm} and Rud et al. \cite{rud2006product}, which can be detected using SDG and information about microservice communication:

\begin{itemize}
    \item Absolute Importance of the Service (AIS) \cite{rud2006product}:  The number of microservices that invoke the current service `a' is the AIS of service `a'. Our analysis can count the number of microservices with an in-degree directed towards service `a' to find AIS (a).
    \item Absolute Dependence of the Service (ADS) \cite{rud2006product}: The number of microservices on which service ‘a’ relies is represented by the ADS of service `a'. We can count the unique microservices with a directed edge from service `a' to find ADS(a).
    \item Cyclic dependency / Services Interdependence in the System (SIY): A service-to-service cycle exists. It could be one, two, or even more services. Even though the requests do not loop because the same endpoints are not being called, this is still a bad practice that can lead to architectural decay. The SIY metric was defined by Rud et al. \cite{rud2006product} as the number of pairs that are dependent on each other. We can look at each pair (a,b) to see if there is a path from service `a' to service `b' and vice versa. The SIY metric is the count of such pairs.
    \item Unbalanced API / Bottleneck Service / Absolute Criticality of the Service (ACS): This anti-pattern appears when a service has a large number of consumers, and the service becomes a single point of failure \cite{nayrolles2013improving, palma2015study}. Rud et al. \cite{rud2006product} defined ACS of service ‘a’ as the product of AIS(a) and ADS(a), which is effectively the product of the number of inbound and outbound microservices of service a. As stated in the definition, we can calculate ACS(a) by multiplying AIS(a) and ADS(a) together (a).
    \item Shared Persistency \cite{taibi2018definition}: This anti-pattern indicates that databases are shared among multiple services. The SDG allows us to determine whether various services access a single database.
    \item API Versioning \cite{taibi2018definition}: APIs should have versions to deal with API changes between versions. Because we have endpoints in our SDG, we can check if they have versioning.
\end{itemize}

\subsection{Software Architecture evolution/Design Rule Change from SDG}

The service dependency graph (SDG) can show the evolution of a system's high-level architecture or design rules. Although software architecture artifacts can provide a service dependency graph, they can quickly become outdated, and MSA commonly decays over time due to architectural debt and code debt. As a result, an SDG generated from runtime logs provides a realistic view of inter-service communication. Furthermore, displaying SDG design rule changes can aid in the verification of new microservice co-relations and the detection of anomalies. It also shows how the new design rule affects real-world traffic.

\subsection{Efficiently scale module using SDG heatmap}

SDG combined with a heat map (number of encounters shown as edge weight) can help scale the system efficiently in resource consumption and cost. For example, the most frequently accessed services in the dependency graph can be identified using the SDG heatmap. Once these services have been identified, developers can make them stateless and deploy multiple copies to increase system throughput. On the other hand, if a service receives little traffic, it is more cost-effective not to scale it. This way, developers can deploy the system more efficiently while reducing costs and resource consumption and increasing system throughput.

\begin{figure*}[!htbp]
  \includegraphics[width=1.3\textwidth, angle=270]{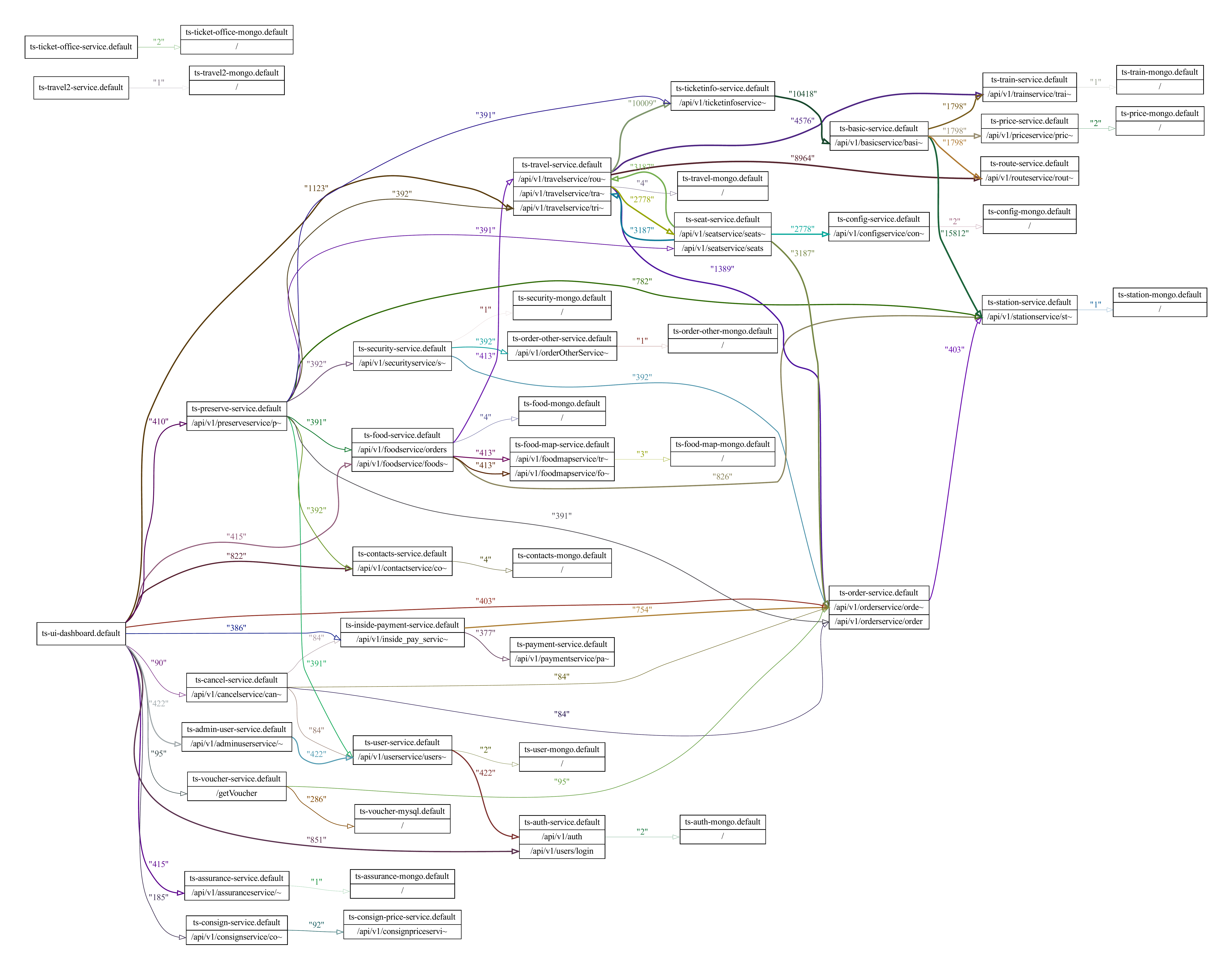}
  \caption{The trainticket-0.1.0 release's Service Dependency Graph (SDG). The graph shows that `ts-travel-service.default' has become one of the system's bottlenecks, while `ts-travel-service.default' and `ts-seat-service.default' have formed a cycle. }
  \label{fig:ttv1}
\end{figure*}

\begin{figure*}[!htbp]
  \includegraphics[width=1.3\textwidth, angle=270]{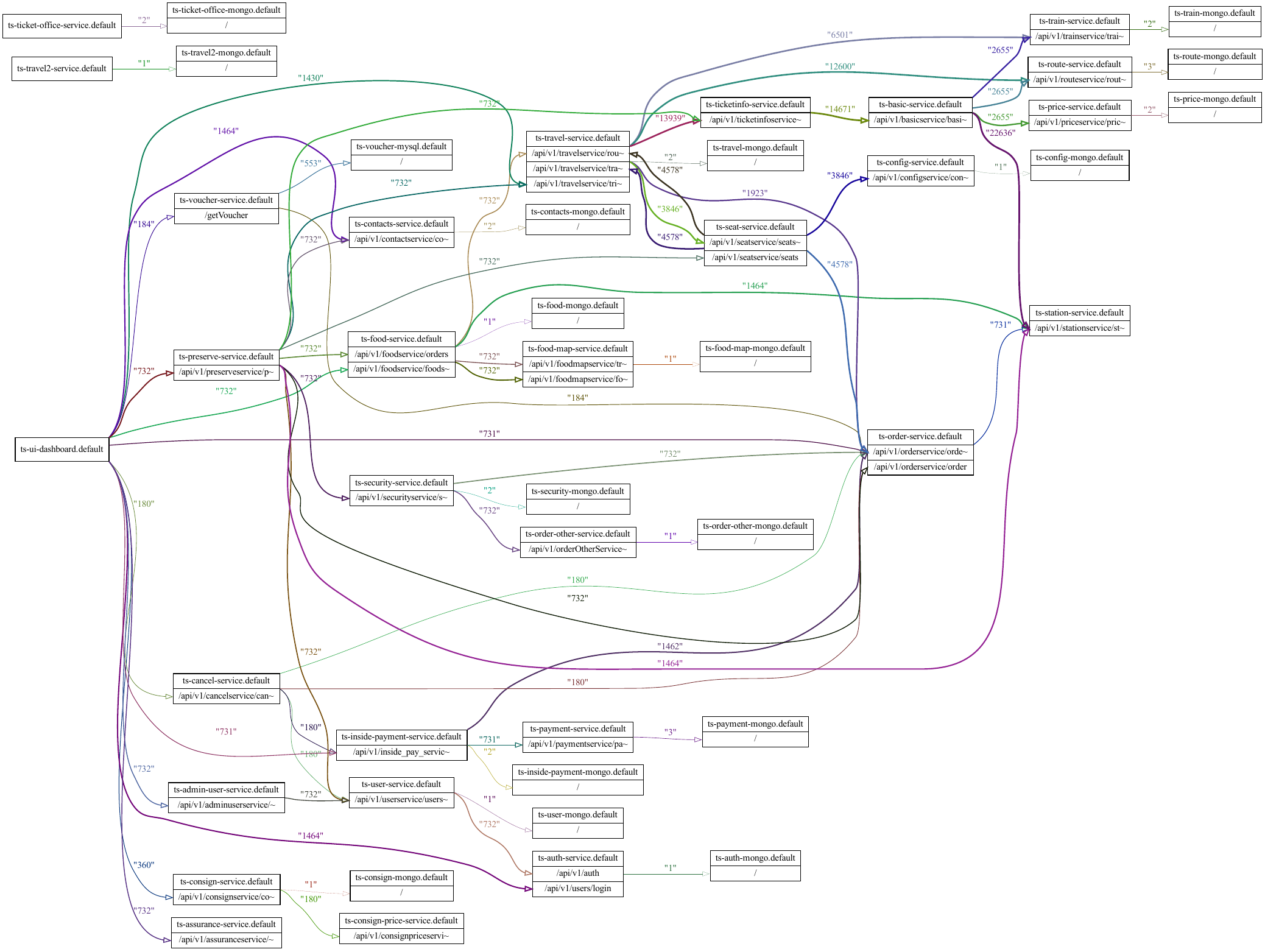}
  \caption{Service Dependency Graph (SDG) of trainticket-0.2.1 release. Similar to v0.1.0, `ts-travel-service.default' has become one of the system's bottlenecks, and a cycle has formed between `ts-travel-service.default' and `ts-seat-service.default'.}
  \label{fig:ttv2}
\end{figure*}

\begin{table*}[ht!]
\begin{tabular}{|l|ccccc|ccccc|}
\hline
\multicolumn{1}{|c|}{\multirow{3}{*}{Service\_Name}} &
  \multicolumn{5}{c|}{Train Ticket v0.2.1} &
  \multicolumn{5}{c|}{Train Ticket v0.1.0} \\ \cline{2-11} 
\multicolumn{1}{|c|}{} &
  \multicolumn{2}{c|}{Service Dependencies} &
  \multicolumn{3}{c|}{\begin{tabular}[c]{@{}c@{}}Criticality and \\ Reliability Metrics\end{tabular}} &
  \multicolumn{2}{c|}{Service Dependencies} &
  \multicolumn{3}{c|}{\begin{tabular}[c]{@{}c@{}}Criticality and \\ Reliability Metrics\end{tabular}} \\ \cline{2-11} 
\multicolumn{1}{|c|}{} &
  \multicolumn{1}{c|}{In-Degrees} &
  \multicolumn{1}{c|}{Out-Degrees} &
  \multicolumn{1}{c|}{\rotV{\begin{tabular}[c]{@{}c@{}}Absolute Importance of the Service \\ (AIS)\end{tabular}}} &
  \multicolumn{1}{c|}{\rotV{\begin{tabular}[c]{@{}c@{}}Absolute Dependence of the Service \\ (ADS)\end{tabular}}} &
  \rotV{\begin{tabular}[c]{@{}c@{}}Absolute Criticality of the Service \\ (ACS)\end{tabular}} &
  \multicolumn{1}{c|}{In-Degrees} &
  \multicolumn{1}{c|}{Out-Degrees} &
  \multicolumn{1}{c|}{\rotV{\begin{tabular}[c]{@{}c@{}}Absolute Importance of the Service \\ (AIS)\end{tabular}}} &
  \multicolumn{1}{c|}{\rotV{\begin{tabular}[c]{@{}c@{}}Absolute Dependence of the Service \\ (ADS)\end{tabular}}} &
  \rotV{\begin{tabular}[c]{@{}c@{}}Absolute Criticality of the Service\\ (ACS)\end{tabular}} \\ \hline
ts-admin-user-service.default &
  \multicolumn{1}{c|}{1} &
  \multicolumn{1}{c|}{1} &
  \multicolumn{1}{c|}{1} &
  \multicolumn{1}{c|}{1} &
  1 &
  \multicolumn{1}{c|}{1} &
  \multicolumn{1}{c|}{1} &
  \multicolumn{1}{c|}{1} &
  \multicolumn{1}{c|}{1} &
  1 \\ \hline
ts-assurance-mongo.default &
  \multicolumn{1}{c|}{} &
  \multicolumn{1}{c|}{} &
  \multicolumn{1}{c|}{} &
  \multicolumn{1}{c|}{} &
   &
  \multicolumn{1}{c|}{1} &
  \multicolumn{1}{c|}{0} &
  \multicolumn{1}{c|}{1} &
  \multicolumn{1}{c|}{0} &
  0 \\ \hline
ts-assurance-service.default &
  \multicolumn{1}{c|}{1} &
  \multicolumn{1}{c|}{0} &
  \multicolumn{1}{c|}{1} &
  \multicolumn{1}{c|}{0} &
  0 &
  \multicolumn{1}{c|}{1} &
  \multicolumn{1}{c|}{1} &
  \multicolumn{1}{c|}{1} &
  \multicolumn{1}{c|}{1} &
  1 \\ \hline
ts-auth-mongo.default &
  \multicolumn{1}{c|}{1} &
  \multicolumn{1}{c|}{0} &
  \multicolumn{1}{c|}{1} &
  \multicolumn{1}{c|}{0} &
  0 &
  \multicolumn{1}{c|}{1} &
  \multicolumn{1}{c|}{0} &
  \multicolumn{1}{c|}{1} &
  \multicolumn{1}{c|}{0} &
  0 \\ \hline
ts-auth-service.default &
  \multicolumn{1}{c|}{2} &
  \multicolumn{1}{c|}{1} &
  \multicolumn{1}{c|}{2} &
  \multicolumn{1}{c|}{1} &
  2 &
  \multicolumn{1}{c|}{2} &
  \multicolumn{1}{c|}{1} &
  \multicolumn{1}{c|}{2} &
  \multicolumn{1}{c|}{1} &
  2 \\ \hline
ts-basic-service.default &
  \multicolumn{1}{c|}{1} &
  \multicolumn{1}{c|}{4} &
  \multicolumn{1}{c|}{1} &
  \multicolumn{1}{c|}{4} &
  4 &
  \multicolumn{1}{c|}{1} &
  \multicolumn{1}{c|}{4} &
  \multicolumn{1}{c|}{1} &
  \multicolumn{1}{c|}{4} &
  4 \\ \hline
ts-cancel-service.default &
  \multicolumn{1}{c|}{1} &
  \multicolumn{1}{c|}{4} &
  \multicolumn{1}{c|}{1} &
  \multicolumn{1}{c|}{4} &
  4 &
  \multicolumn{1}{c|}{1} &
  \multicolumn{1}{c|}{4} &
  \multicolumn{1}{c|}{1} &
  \multicolumn{1}{c|}{4} &
  4 \\ \hline
ts-config-mongo.default &
  \multicolumn{1}{c|}{1} &
  \multicolumn{1}{c|}{0} &
  \multicolumn{1}{c|}{1} &
  \multicolumn{1}{c|}{0} &
  0 &
  \multicolumn{1}{c|}{1} &
  \multicolumn{1}{c|}{0} &
  \multicolumn{1}{c|}{1} &
  \multicolumn{1}{c|}{0} &
  0 \\ \hline
ts-config-service.default &
  \multicolumn{1}{c|}{1} &
  \multicolumn{1}{c|}{1} &
  \multicolumn{1}{c|}{1} &
  \multicolumn{1}{c|}{1} &
  1 &
  \multicolumn{1}{c|}{1} &
  \multicolumn{1}{c|}{1} &
  \multicolumn{1}{c|}{1} &
  \multicolumn{1}{c|}{1} &
  1 \\ \hline
ts-consign-mongo.default &
  \multicolumn{1}{c|}{1} &
  \multicolumn{1}{c|}{0} &
  \multicolumn{1}{c|}{1} &
  \multicolumn{1}{c|}{0} &
  0 &
  \multicolumn{1}{c|}{} &
  \multicolumn{1}{c|}{} &
  \multicolumn{1}{c|}{} &
  \multicolumn{1}{c|}{} &
   \\ \hline
ts-consign-price-service.default &
  \multicolumn{1}{c|}{1} &
  \multicolumn{1}{c|}{0} &
  \multicolumn{1}{c|}{1} &
  \multicolumn{1}{c|}{0} &
  0 &
  \multicolumn{1}{c|}{1} &
  \multicolumn{1}{c|}{0} &
  \multicolumn{1}{c|}{1} &
  \multicolumn{1}{c|}{0} &
  0 \\ \hline
ts-consign-service.default &
  \multicolumn{1}{c|}{1} &
  \multicolumn{1}{c|}{2} &
  \multicolumn{1}{c|}{1} &
  \multicolumn{1}{c|}{2} &
  2 &
  \multicolumn{1}{c|}{1} &
  \multicolumn{1}{c|}{1} &
  \multicolumn{1}{c|}{1} &
  \multicolumn{1}{c|}{1} &
  1 \\ \hline
ts-contacts-mongo.default &
  \multicolumn{1}{c|}{1} &
  \multicolumn{1}{c|}{0} &
  \multicolumn{1}{c|}{1} &
  \multicolumn{1}{c|}{0} &
  0 &
  \multicolumn{1}{c|}{1} &
  \multicolumn{1}{c|}{0} &
  \multicolumn{1}{c|}{1} &
  \multicolumn{1}{c|}{0} &
  0 \\ \hline
ts-contacts-service.default &
  \multicolumn{1}{c|}{2} &
  \multicolumn{1}{c|}{1} &
  \multicolumn{1}{c|}{2} &
  \multicolumn{1}{c|}{1} &
  2 &
  \multicolumn{1}{c|}{2} &
  \multicolumn{1}{c|}{1} &
  \multicolumn{1}{c|}{2} &
  \multicolumn{1}{c|}{1} &
  2 \\ \hline
ts-food-map-mongo.default &
  \multicolumn{1}{c|}{1} &
  \multicolumn{1}{c|}{0} &
  \multicolumn{1}{c|}{1} &
  \multicolumn{1}{c|}{0} &
  0 &
  \multicolumn{1}{c|}{1} &
  \multicolumn{1}{c|}{0} &
  \multicolumn{1}{c|}{1} &
  \multicolumn{1}{c|}{0} &
  0 \\ \hline
ts-food-map-service.default &
  \multicolumn{1}{c|}{2} &
  \multicolumn{1}{c|}{1} &
  \multicolumn{1}{c|}{2} &
  \multicolumn{1}{c|}{1} &
  2 &
  \multicolumn{1}{c|}{2} &
  \multicolumn{1}{c|}{1} &
  \multicolumn{1}{c|}{2} &
  \multicolumn{1}{c|}{1} &
  2 \\ \hline
ts-food-mongo.default &
  \multicolumn{1}{c|}{1} &
  \multicolumn{1}{c|}{0} &
  \multicolumn{1}{c|}{1} &
  \multicolumn{1}{c|}{0} &
  0 &
  \multicolumn{1}{c|}{1} &
  \multicolumn{1}{c|}{0} &
  \multicolumn{1}{c|}{1} &
  \multicolumn{1}{c|}{0} &
  0 \\ \hline
ts-food-service.default &
  \multicolumn{1}{c|}{2} &
  \multicolumn{1}{c|}{5} &
  \multicolumn{1}{c|}{2} &
  \multicolumn{1}{c|}{5} &
  10 &
  \multicolumn{1}{c|}{2} &
  \multicolumn{1}{c|}{5} &
  \multicolumn{1}{c|}{2} &
  \multicolumn{1}{c|}{5} &
  10 \\ \hline
ts-inside-payment-mongo.default &
  \multicolumn{1}{c|}{1} &
  \multicolumn{1}{c|}{0} &
  \multicolumn{1}{c|}{1} &
  \multicolumn{1}{c|}{0} &
  0 &
  \multicolumn{1}{c|}{} &
  \multicolumn{1}{c|}{} &
  \multicolumn{1}{c|}{} &
  \multicolumn{1}{c|}{} &
   \\ \hline
ts-inside-payment-service.default &
  \multicolumn{1}{c|}{2} &
  \multicolumn{1}{c|}{3} &
  \multicolumn{1}{c|}{2} &
  \multicolumn{1}{c|}{3} &
  6 &
  \multicolumn{1}{c|}{2} &
  \multicolumn{1}{c|}{2} &
  \multicolumn{1}{c|}{2} &
  \multicolumn{1}{c|}{2} &
  4 \\ \hline
ts-order-other-mongo.default &
  \multicolumn{1}{c|}{1} &
  \multicolumn{1}{c|}{0} &
  \multicolumn{1}{c|}{1} &
  \multicolumn{1}{c|}{0} &
  0 &
  \multicolumn{1}{c|}{1} &
  \multicolumn{1}{c|}{0} &
  \multicolumn{1}{c|}{1} &
  \multicolumn{1}{c|}{0} &
  0 \\ \hline
ts-order-other-service.default &
  \multicolumn{1}{c|}{1} &
  \multicolumn{1}{c|}{1} &
  \multicolumn{1}{c|}{1} &
  \multicolumn{1}{c|}{1} &
  1 &
  \multicolumn{1}{c|}{1} &
  \multicolumn{1}{c|}{1} &
  \multicolumn{1}{c|}{1} &
  \multicolumn{1}{c|}{1} &
  1 \\ \hline
ts-order-service.default &
  \multicolumn{1}{c|}{9} &
  \multicolumn{1}{c|}{1} &
  \multicolumn{1}{c|}{9} &
  \multicolumn{1}{c|}{1} &
  9 &
  \multicolumn{1}{c|}{9} &
  \multicolumn{1}{c|}{1} &
  \multicolumn{1}{c|}{9} &
  \multicolumn{1}{c|}{1} &
  9 \\ \hline
ts-payment-mongo.default &
  \multicolumn{1}{c|}{1} &
  \multicolumn{1}{c|}{0} &
  \multicolumn{1}{c|}{1} &
  \multicolumn{1}{c|}{0} &
  0 &
  \multicolumn{1}{c|}{} &
  \multicolumn{1}{c|}{} &
  \multicolumn{1}{c|}{} &
  \multicolumn{1}{c|}{} &
   \\ \hline
ts-payment-service.default &
  \multicolumn{1}{c|}{1} &
  \multicolumn{1}{c|}{1} &
  \multicolumn{1}{c|}{1} &
  \multicolumn{1}{c|}{1} &
  1 &
  \multicolumn{1}{c|}{1} &
  \multicolumn{1}{c|}{0} &
  \multicolumn{1}{c|}{1} &
  \multicolumn{1}{c|}{0} &
  0 \\ \hline
ts-preserve-service.default &
  \multicolumn{1}{c|}{1} &
  \multicolumn{1}{c|}{9} &
  \multicolumn{1}{c|}{1} &
  \multicolumn{1}{c|}{9} &
  9 &
  \multicolumn{1}{c|}{1} &
  \multicolumn{1}{c|}{9} &
  \multicolumn{1}{c|}{1} &
  \multicolumn{1}{c|}{9} &
  9 \\ \hline
ts-price-mongo.default &
  \multicolumn{1}{c|}{1} &
  \multicolumn{1}{c|}{0} &
  \multicolumn{1}{c|}{1} &
  \multicolumn{1}{c|}{0} &
  0 &
  \multicolumn{1}{c|}{1} &
  \multicolumn{1}{c|}{0} &
  \multicolumn{1}{c|}{1} &
  \multicolumn{1}{c|}{0} &
  0 \\ \hline
ts-price-service.default &
  \multicolumn{1}{c|}{1} &
  \multicolumn{1}{c|}{1} &
  \multicolumn{1}{c|}{1} &
  \multicolumn{1}{c|}{1} &
  1 &
  \multicolumn{1}{c|}{1} &
  \multicolumn{1}{c|}{1} &
  \multicolumn{1}{c|}{1} &
  \multicolumn{1}{c|}{1} &
  1 \\ \hline
ts-route-mongo.default &
  \multicolumn{1}{c|}{1} &
  \multicolumn{1}{c|}{0} &
  \multicolumn{1}{c|}{1} &
  \multicolumn{1}{c|}{0} &
  0 &
  \multicolumn{1}{c|}{} &
  \multicolumn{1}{c|}{} &
  \multicolumn{1}{c|}{} &
  \multicolumn{1}{c|}{} &
   \\ \hline
ts-route-service.default &
  \multicolumn{1}{c|}{2} &
  \multicolumn{1}{c|}{1} &
  \multicolumn{1}{c|}{2} &
  \multicolumn{1}{c|}{1} &
  2 &
  \multicolumn{1}{c|}{2} &
  \multicolumn{1}{c|}{0} &
  \multicolumn{1}{c|}{2} &
  \multicolumn{1}{c|}{0} &
  0 \\ \hline
ts-seat-service.default &
  \multicolumn{1}{c|}{2} &
  \multicolumn{1}{c|}{4} &
  \multicolumn{1}{c|}{2} &
  \multicolumn{1}{c|}{4} &
  8 &
  \multicolumn{1}{c|}{2} &
  \multicolumn{1}{c|}{4} &
  \multicolumn{1}{c|}{2} &
  \multicolumn{1}{c|}{4} &
  8 \\ \hline
ts-security-mongo.default &
  \multicolumn{1}{c|}{1} &
  \multicolumn{1}{c|}{0} &
  \multicolumn{1}{c|}{1} &
  \multicolumn{1}{c|}{0} &
  0 &
  \multicolumn{1}{c|}{1} &
  \multicolumn{1}{c|}{0} &
  \multicolumn{1}{c|}{1} &
  \multicolumn{1}{c|}{0} &
  0 \\ \hline
ts-security-service.default &
  \multicolumn{1}{c|}{1} &
  \multicolumn{1}{c|}{3} &
  \multicolumn{1}{c|}{1} &
  \multicolumn{1}{c|}{3} &
  3 &
  \multicolumn{1}{c|}{1} &
  \multicolumn{1}{c|}{3} &
  \multicolumn{1}{c|}{1} &
  \multicolumn{1}{c|}{3} &
  3 \\ \hline
ts-station-mongo.default &
  \multicolumn{1}{c|}{} &
  \multicolumn{1}{c|}{} &
  \multicolumn{1}{c|}{} &
  \multicolumn{1}{c|}{} &
   &
  \multicolumn{1}{c|}{1} &
  \multicolumn{1}{c|}{0} &
  \multicolumn{1}{c|}{1} &
  \multicolumn{1}{c|}{0} &
  0 \\ \hline
ts-station-service.default &
  \multicolumn{1}{c|}{4} &
  \multicolumn{1}{c|}{0} &
  \multicolumn{1}{c|}{4} &
  \multicolumn{1}{c|}{0} &
  0 &
  \multicolumn{1}{c|}{4} &
  \multicolumn{1}{c|}{1} &
  \multicolumn{1}{c|}{4} &
  \multicolumn{1}{c|}{1} &
  4 \\ \hline
ts-ticket-office-mongo.default &
  \multicolumn{1}{c|}{1} &
  \multicolumn{1}{c|}{0} &
  \multicolumn{1}{c|}{1} &
  \multicolumn{1}{c|}{0} &
  0 &
  \multicolumn{1}{c|}{1} &
  \multicolumn{1}{c|}{0} &
  \multicolumn{1}{c|}{1} &
  \multicolumn{1}{c|}{0} &
  0 \\ \hline
ts-ticket-office-service.default &
  \multicolumn{1}{c|}{0} &
  \multicolumn{1}{c|}{1} &
  \multicolumn{1}{c|}{0} &
  \multicolumn{1}{c|}{1} &
  0 &
  \multicolumn{1}{c|}{0} &
  \multicolumn{1}{c|}{1} &
  \multicolumn{1}{c|}{0} &
  \multicolumn{1}{c|}{1} &
  0 \\ \hline
ts-ticketinfo-service.default &
  \multicolumn{1}{c|}{2} &
  \multicolumn{1}{c|}{1} &
  \multicolumn{1}{c|}{2} &
  \multicolumn{1}{c|}{1} &
  2 &
  \multicolumn{1}{c|}{2} &
  \multicolumn{1}{c|}{1} &
  \multicolumn{1}{c|}{2} &
  \multicolumn{1}{c|}{1} &
  2 \\ \hline
ts-train-mongo.default &
  \multicolumn{1}{c|}{1} &
  \multicolumn{1}{c|}{0} &
  \multicolumn{1}{c|}{1} &
  \multicolumn{1}{c|}{0} &
  0 &
  \multicolumn{1}{c|}{1} &
  \multicolumn{1}{c|}{0} &
  \multicolumn{1}{c|}{1} &
  \multicolumn{1}{c|}{0} &
  0 \\ \hline
ts-train-service.default &
  \multicolumn{1}{c|}{2} &
  \multicolumn{1}{c|}{1} &
  \multicolumn{1}{c|}{2} &
  \multicolumn{1}{c|}{1} &
  2 &
  \multicolumn{1}{c|}{2} &
  \multicolumn{1}{c|}{1} &
  \multicolumn{1}{c|}{2} &
  \multicolumn{1}{c|}{1} &
  2 \\ \hline
ts-travel-mongo.default &
  \multicolumn{1}{c|}{1} &
  \multicolumn{1}{c|}{0} &
  \multicolumn{1}{c|}{1} &
  \multicolumn{1}{c|}{0} &
  0 &
  \multicolumn{1}{c|}{1} &
  \multicolumn{1}{c|}{0} &
  \multicolumn{1}{c|}{1} &
  \multicolumn{1}{c|}{0} &
  0 \\ \hline
ts-travel-service.default &
  \multicolumn{1}{c|}{5} &
  \multicolumn{1}{c|}{6} &
  \multicolumn{1}{c|}{5} &
  \multicolumn{1}{c|}{6} &
  30 &
  \multicolumn{1}{c|}{5} &
  \multicolumn{1}{c|}{6} &
  \multicolumn{1}{c|}{5} &
  \multicolumn{1}{c|}{6} &
  30 \\ \hline
ts-travel2-mongo.default &
  \multicolumn{1}{c|}{1} &
  \multicolumn{1}{c|}{0} &
  \multicolumn{1}{c|}{1} &
  \multicolumn{1}{c|}{0} &
  0 &
  \multicolumn{1}{c|}{1} &
  \multicolumn{1}{c|}{0} &
  \multicolumn{1}{c|}{1} &
  \multicolumn{1}{c|}{0} &
  0 \\ \hline
ts-travel2-service.default &
  \multicolumn{1}{c|}{0} &
  \multicolumn{1}{c|}{1} &
  \multicolumn{1}{c|}{0} &
  \multicolumn{1}{c|}{1} &
  0 &
  \multicolumn{1}{c|}{0} &
  \multicolumn{1}{c|}{1} &
  \multicolumn{1}{c|}{0} &
  \multicolumn{1}{c|}{1} &
  0 \\ \hline
ts-ui-dashboard.default &
  \multicolumn{1}{c|}{0} &
  \multicolumn{1}{c|}{12} &
  \multicolumn{1}{c|}{0} &
  \multicolumn{1}{c|}{12} &
  0 &
  \multicolumn{1}{c|}{0} &
  \multicolumn{1}{c|}{12} &
  \multicolumn{1}{c|}{0} &
  \multicolumn{1}{c|}{12} &
  0 \\ \hline
ts-user-mongo.default &
  \multicolumn{1}{c|}{1} &
  \multicolumn{1}{c|}{0} &
  \multicolumn{1}{c|}{1} &
  \multicolumn{1}{c|}{0} &
  0 &
  \multicolumn{1}{c|}{1} &
  \multicolumn{1}{c|}{0} &
  \multicolumn{1}{c|}{1} &
  \multicolumn{1}{c|}{0} &
  0 \\ \hline
ts-user-service.default &
  \multicolumn{1}{c|}{3} &
  \multicolumn{1}{c|}{2} &
  \multicolumn{1}{c|}{3} &
  \multicolumn{1}{c|}{2} &
  6 &
  \multicolumn{1}{c|}{3} &
  \multicolumn{1}{c|}{2} &
  \multicolumn{1}{c|}{3} &
  \multicolumn{1}{c|}{2} &
  6 \\ \hline
ts-voucher-mysql.default &
  \multicolumn{1}{c|}{1} &
  \multicolumn{1}{c|}{0} &
  \multicolumn{1}{c|}{1} &
  \multicolumn{1}{c|}{0} &
  0 &
  \multicolumn{1}{c|}{1} &
  \multicolumn{1}{c|}{0} &
  \multicolumn{1}{c|}{1} &
  \multicolumn{1}{c|}{0} &
  0 \\ \hline
ts-voucher-service.default &
  \multicolumn{1}{c|}{1} &
  \multicolumn{1}{c|}{2} &
  \multicolumn{1}{c|}{1} &
  \multicolumn{1}{c|}{2} &
  2 &
  \multicolumn{1}{c|}{1} &
  \multicolumn{1}{c|}{2} &
  \multicolumn{1}{c|}{1} &
  \multicolumn{1}{c|}{2} &
  2 \\ \hline
\end{tabular}
\caption{Comparison of antipattern metrics that are calculated automatically.}
\label{tab:metrics}
\end{table*}

\section{Case Study}  \label{case-study}

We developed the prototype istio-log-parser tool \footnote{https://github.com/the-redback/istio-log-parsing}, which takes an access log as input and generates a Service Dependency Graph with heatmap. It also calculates anti-pattern metrics and examines the system for any cycles. It generates a CSV file with anti-pattern metrics calculated, and the cycles are displayed in the program's stdout. To create the graph, we used Graphviz \footnote{https://graphviz.org/}, which generates a dot file, which we then converted to pdf format using the dot library \cite{gansner2000open}.
We used a microservice benchmark called TrainTicket \cite{zhou2018benchmarking} to test the prototype. The microservice's most recent release has around 270,000 lines of code. To test the prototype, we utilized both the v0.2.1 and v0.1.0 releases. TrainTicket v0.1.0 has 41 modules, while v0.2.1 has 42, and both use a reverse proxy based on Nginx to route requests to modules. The system was deployed on Kubernetes, along with Istio. The Istio was set up in such a way that it generates and outputs a JSON-formatted access log. We installed Kubernetes in a machine with 32GB of RAM and an Intel i9-8950HK processor using kind (Kubernetes-in-docker)\footnote{https://kind.sigs.k8s.io/}. There were 6 CPU cores and 12 threads in the system. Kubernetes' resource limit was set at 22GB of RAM and 8 threads of CPU.

We used the benchmarking tool PPTAM \cite{avritzer2019pptam} to simulate a real-time user. It had five scenarios to simulate, and five users were making requests in these five scenarios at the same time. The test lasted 30 minutes and resulted in a total of ~10,000 requests being sent to the server. After the simulation was finished, we collected each service's Istio proxy logs and ran our prototype tool on them. Both v0.2.1 and v0.1.0 went through a similar simulation process.
 
\subsection{Anti-Patterns}
Figure \ref{fig:ttv1} is the generated SDG with heatmap of trainticket-v0.1.0 and Figure \ref{fig:ttv2} is the SDG of v0.2.1. Table \ref{tab:metrics} shows a comparison of the automatically calculated anti-pattern metrics.  From this available information we can find the followings about anti-pattern:
\begin{itemize}
    \item Absolute Importance of the Service (AIS):  We can see from Table-\ref{tab:metrics} that `ts-order-service.default' has the most AIS (9). In addition, few services with 0 AIS indicate that they have no consumers or clients. Although this is not true for `ts-ui-dashboard.default', because it is a reverse proxy, there are no inbound services, and the only consumers are requests from outside the cluster/microservice, which is referred to as ingress.
    \item Absolute Dependence of the Service (ADS): The outdegree of `ts-ui-dasshboard.default' is twelve. As a result, it has the highest ADS in the system. ts-preserve-service.default has the second-highest ADS at nine.
    \item Cyclic dependency / Services Interdependence in the System (SIY): Both versions have a cycle between the services `ts-travel-service.default' and `ts-seat-service.default'. We can verify that the prototype's findings are correct by looking at Figure-\ref{fig:ttv1} and Figure-\ref{fig:ttv2}.
    \item Unbalanced API / Bottleneck Service / Absolute Criticality of the Service (ACS): We can see from Table-\ref{tab:metrics} that the ACS value for 	`ts-travel-service.default' is the highest. It applies to both versions v0.1.0 and v0.2.1. When we look at the SDG, we can see how this service has become a system bottleneck and a single point of failure. There are five services that `ts-travel-service.default' is a client of, and six services that `ts-travel-service.default' accesses. The ACS value for 'ts-food-service.default' is the second highest.
    \item Shared Persistency: It is clear from the SDGs (Figure-\ref{fig:ttv1} and Figure-\ref{fig:ttv2}) that no MongoDB is shared by multiple services.
    \item API Versioning: We can see from both SDGs (Figure-\ref{fig:ttv1} and Figure-\ref{fig:ttv2}) that each endpoint has 1/api/v1' as a prefix to the endpoint. This indicates that API versioning is in effect.
\end{itemize}

\subsection{Evolution of MSA}

The SDGs in versions 0.2.1 and 0.1.0 have the same number of services, with the exception that version 0.1.0 lacks a few MongoDB connections. Despite the fact that version 0.2.1 includes an additional module called 'ts-delivery-service,' it is not visible in the graph. This is because the PPTAM testing artifacts are older and do not cover the scenario of newly added business logic/service. Although this is a limitation of dynamic analysis, it does provide a scenario to verify that, due to API versioning, older tests are still functional without introducing new business flows.

\subsection{Efficiently scale module}

The services with the most inbound edges should be scaled first, as they are the most important for others to communicate with. In this case, it's necessary to scale `ts-order-service. default' first. `ts-travel-service.default' is the second service that needs to be scaled, but it has become the system's bottleneck with the highest absolute critical ACS value. Therefore, developers of such microservices must first reduce tanglement before scaling the service. The following microservices to scale are those with the highest ADS values and the most outbound edges. Because these services rely on various other services, the response time for each request to this service may increase, resulting in a queue for its users. Nevertheless, it is still necessary to distribute it in order to handle more users.

\section{Conclusion} \label{conclusion}

This paper presents broad advancement to the microservice community, combining the perspective of dynamic analysis with quality assessment applicable to system evolution. It addresses current challenges in microservices. Challenges that the static analysis did not yet overcome due to decentralization and possible language heterogeneity across microservice codebases. It demonstrates the feasibility of anomaly detection on established telemetry tools and combines quality assurance knowledge established in service-oriented systems. We demonstrated the feasibility of the approach on a complex system benchmark with 42 microservices, showed detection of anti-patterns and metrics and presented a use case when applied to the evolving system. The results can give practitioners an automated tool to detect anomalies in non-centrally evolving systems and the ability to prioritize identified tasks.

In future work, we anticipate integrating business process analysis and enabling inter-weaving with anti-patterns. We also aim to study other microservice-relevant patterns that could be detected through dynamic analysis, such us \texttt{\url{microservice-api-patterns.org}}. We aim to integrate our detection and metrics into established tooling. Furthermore, despite the potential, we plan to compare it with static analysis.

\section*{Acknowledgment}
This material is based upon work supported by the National Science Foundation under Grant No. 1854049, grant from Red Hat Research, and Ulla Tuominen (Shapit).

\bibliographystyle{IEEEtran}
\bibliography{access}

\end{document}